\documentclass[11pt,twoside]{article}
\usepackage{asp2010}

\newcommand{\be}{\begin{equation}}

\newcommand{\ee}{\end{equation}}

\def\al{Alfv\'en\ }

\def\apjl{Astrophys. J. Lett.}                
\def\aap{Astron. \& Astrophys.}                

\begin{document}
\resetcounters
\markboth{Lin, Ng, and Bhattacharjee}{Current Sheet Statistics in 3D MHD Simulations}

\title{Current Sheet Statistics in Three-Dimensional Simulations of Coronal Heating}
\author{L. Lin$^1$, C.S. Ng$^2$, and A. Bhattacharjee$^{1,3}$}
\affil{$^1$Space Science Center, University of New Hampshire, Durham, NH 03824}
\affil{$^2$Geophysical Institute, University of Alaska Fairbanks, Fairbanks, AK
99775}
\affil{$^3$Princeton Plasma Physics Laboratory, Princeton, NJ 08543}

\begin{abstract}
In a recent numerical study [Ng et al., Astrophys. J. {\bf 747}, 109, 2012], with a
three-dimensional model of coronal heating using reduced
magnetohydrodynamics (RMHD), we have obtained scaling results of heating rate
versus Lundquist number based on a series of runs in which random photospheric
motions are imposed for hundreds to thousands of \al time in order to
obtain converged statistical values. The heating rate found in these
simulations saturate to a level that is independent of the Lundquist number. This
scaling result was also supported by an analysis with the assumption of the
Sweet-Parker scaling of the current sheets, as well as how the width, length
and number of current sheets scale with Lundquist number. In order to test
these assumptions, we have implemented an automated routine to analyze
thousands of current sheets in these simulations and return statistical
scalings for these quantities. It is found that the Sweet-Parker scaling
is justified. However, some discrepancies are also found and require further
study. 		
\end{abstract}

\section{Introduction} 
Within the framework of the parker model of coronal heating \citep{Parker1972},
a recent analysis \citep{nb2008} in two dimensions (2D) demonstrated  that when
coherence times ($\tau_{c}$) of the imposed photospheric turbulence are much
smaller than characteristic resistive time-scales ($\tau_{R}$), the Ohmic
dissipation scales independently of resistivity. While their initial  2D RMHD
treatment precluded non-linear effects such as instabilities and/or magnetic
reconnection, they further invoked a simple analytical argument
demonstrated that even with these non-linear effects, which would limit
the growth of $B_{\perp}$ (the component of the magnetic field perpendicular to
a guide field $B_z$),  the
insensitivity to resistivity is still true for small enough $\tau_{c}$. 
This latter hypothesis was
subsequently confirmed \citep{nlb2012} by means of 3D RMHD
simulations of the Parker model which spanned three orders of magnitude 
in Lundquist number. 

The scalings derived in these studies  depend critically on the assumption
that the classical 2D steady-state Sweet-Parker scaling for magnetic
reconnection holds in 3D simulations where extended current sheets form due to
random boundary driving. In order to empirically substantiate this assumption,
and to look into the nature of stochastically driven
magentofluids in more details, we have carried out a systematic analysis of current sheets
formed in our simulations. 

We report here a simple algorithm to identify and characterize individual
current layers. 
Statistics are accumulated for current sheet parameters. 
In Section 2, we develop and review the motivating
analysis.  Section 3 describes our simple algorithm. Section 4 reports our
findings and outstanding issues with our analysis.

\begin{figure}[t]
\centering
\includegraphics[scale=0.30]{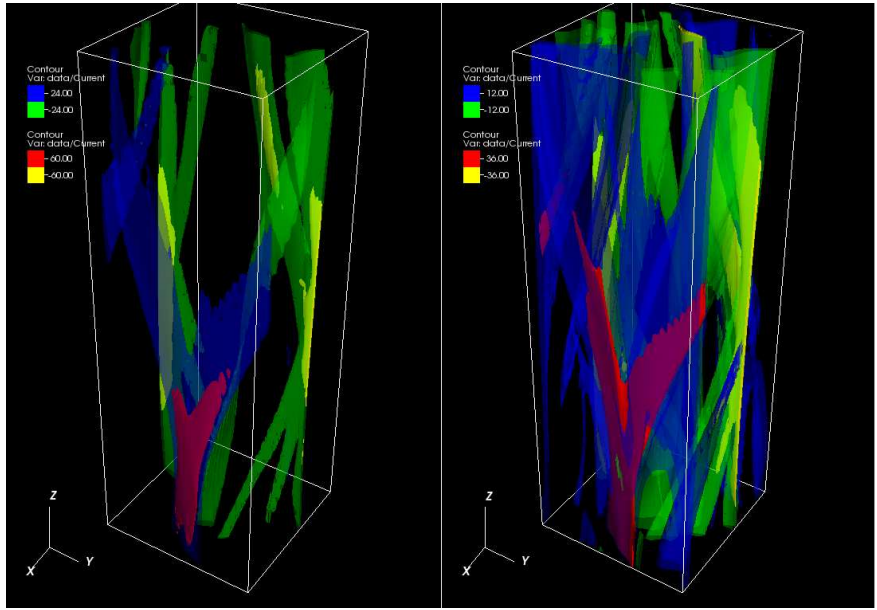}
\caption{3D iso-surfaces of current sheets.} 
\label{cs}

\end{figure}

\section{Coronal Heating Scaling Analysis} 
\label{1scal}
In the parker model, a solar coronal loop
  is treated as a straight ideal plasma column, 
bounded by two perfectly conducting end-plates representing the photosphere.  
Initially, there is a uniform magnetic field along the $z$  direction.  
The footpoints of the magnetic field on the photosphere are frozen (line-tied),
subjected to slow, 
random motion $\phi(z=0,t)$ and $\phi(z=L,t)$ that deform the magnetic field. 

The footpoint motion is assumed to take place on a time scale much longer 
  than the characteristic time for \al wave propagation between $z = 0$  and $z = L$, 
so that the plasma can be assumed to be in quasi static equilibrium nearly everywhere, 
if such equilibrium exists, during this random evolution.  
For a given equilibrium, 
a footpoint mapping can be defined by following field lines from one plate to the other.  
Since the plasma is assumed to obey the ideal MHD equations, 
the magnetic field lines are frozen in the plasma and cannot be broken during the twisting process.  
Therefore, the footpoint mapping must be continuous for smooth footpoint motion.  
\cite{Parker1972} claims that if a sequence of random footpoint motion 
  renders the mapping sufficiently complicated, 
there will be no smooth equilibrium for the plasma to relax to, 
and tangential discontinuities (or current sheets) of the magnetic field must develop.
Parker treated the corona as ideal given that the Lundquist number of the
corona is estimated to be of the order of $10^{13}$. Being of ultimately
finite resistivity however, it is suggested that ohmic dissipation in these
current sheets can significantly account for heating coronal plasma. 

Resolving current sheets at realistic values of Lundquist numbers remains well
beyond the reach of current computational capabilities with direct numerical
MHD simulations. Useful scaling studies, however, have been carried out by numerous
investigators (cf. \citealt{nlb2012} for references). Our current numerical study is
motivated by scaling analysis for coronal heating that, unlike previous
derivations of the coronal heating rate, considers the effects of random
footpoint motion. We quickly summarize the main arguments here (cf. \citealt{nb2008} and \citealt{nlb2012} for details). 

In well resolved direct time-dependent, 3D RMHD numerical simulations of the Parker model, 
on average, at any given time, there will be $N$ current sheets with characteristic width ($\lambda$)
and length ($\Delta$), which span the length of the simulation box ($L$). The
characteristic time over which energy is built-up by random photospheric
motions and subsequently released is $\tau_{E}$, and therfore the average
heating rate due to ohmic dissipation can be written as:

\begin{equation} 
 \bar{W} \sim \eta N \Delta L \frac{\bar{B}^2_\perp}{\lambda} \sim 
   \frac{\bar{B}^2_\perp L L^2_\perp}{\tau_E},
\label{wavg}\end{equation} 
where $\eta$ is the plasma resistivity and $\bar{B}^2_\perp$ is the average magnetic field.  
An expression for the average perpendicular magnetic field can be estimated
considering that over a duration of ${\tau_{E}}$, photospheric footpoint motions of
average velocity $v_{p}$, if assumed constant, would deform the guide field
$B_z$ and produce  $\bar{B}_\perp$ up to a level of 

\begin{equation}
  \bar{B}_\perp \sim B_z \frac{v_p \tau_E}{L} 
  \sim \left[ \left( \frac{B_z v_p}{L N}\right)^2 \frac{L^4_\perp}{w \eta} \right]^{1/3} ,
\label{Bperp}\end{equation}
where, for the latter expression, we have solved for ${\tau_{E}}$ in Equation
(\ref{wavg}) using the Sweet-Parker current sheet scaling given by
$\lambda/\Delta \sim S_{\perp}^{1/2}$. We define the Lundquist number here as
$S_{\perp}\equiv w B_{\perp}/\eta$, with $w=v_p\tau_c$ being the typical transverse 
length scale of the magnetic field. Together,
Equations (\ref{wavg}) and (\ref{Bperp}) yield the following expression for the
average heating rate: 

\begin{equation} 
   \bar{W} \sim  \left( \frac{L^{10}_\perp B_z^5 v_p^5}{L^2 N^2 w \eta}\right)^{1/3}.
\label{wbar-long}\end{equation}

Evidently, when one extrapolates to the collisionless coronal limit, the
heating rate predicted here becomes un-physically large. If we rewrite however
the expression for the perpendicular magnetic field production considering
the turbulent motions in the photosphere to have a random walk nature: 
$\bar{B}_\perp \sim B_{z}v_{p}(\tau_c \tau_E)^{1/2}/L$, 
an average heating rate can be estimated as: 

\begin{equation} 
   \bar{W} \sim  \frac{L^2_\perp}{L} B_z^2 v_p^2 \tau_c .
\label{wbar-ran}\end{equation}

Equation (\ref{wbar-ran}) is manifestly independent of resistivity, and holds
when $\tau_c < \tau_E$, i.e. when the effects of random motion become
important. In \cite{nlb2012} we have studied the transition of the heating
rate into this regime with numerical simulations spanning three orders of
magnitude. In the regime where $\tau_c > \tau_E$, we find good agreement with
a previous study \citep{ls1994}, with the heating rate scales as  
$\bar{W}\propto{\eta^{1/3}}$.
While in the high Lundquist number regime where
$\tau_c < \tau_E$, we recover an $\eta$ independent behavior. The reader is
referred to \cite{nlb2012} for an in depth discussion of these results. Here
we focus on an ancillary study that addresses two key assumptions made in
arriving at the scaling relations here: $(1)$ The number of current sheets $N$
is essentially independent of Lundquist number; and $(2)$ The Sweet-Parker
scaling, which derives from a 2D quasi-steady theory, is applicable more
generally in our 3D line-tied model with driving applied at the boundaries.
We assess these two assumptions by means of a straightforward algorithm.

\section{Current Sheet Identification and Fitting}
\label{2cif}
Our MHD simulations employ the reduced MHD 
equations and are solved using a standard algorithm (described in \citealt{nlb2012},
and have been recently accelerated via GPUs with NVidia
CUDA as shown in \citealt{LNB2012}). 
Our current goal is to identify and characterize individual current layers forming in 
time-dependent 3D simulations of a coronal loop
driven at the line-tied boundaries. Figure~\ref{cs} shows iso-surfaces of current
density for a snapshot of one of our simulation runs (with $S_{\perp}
\sim5000$). As a starting point, we examine only the 2D cross section at $z=L/2$
 of our simulation domain.

This task is formidable for the following reasons: $(1)$ Given the stochastic 
nature of the imposed photospheric boundary driving, current sheet orientations are random. 
$(2)$ We are dealing with tens of thousands of individual instances of
current sheets forming during steady state evolution of the Parker model, for
which we have data cubes saved at a prescribed cadence.
$(3)$ We use periodic boundary conditions in which current sheets often
traverse the edges. $(4)$ In three dimensions, current sheets appear to branch
out, so a structure appearing as a single current layer in one specific
cross-section of the loop might appear as several in a different cross-section
at a location further along the loop, possibly with different properties. 
Figure~\ref{cs} attests to each of these issues. 

\begin{figure}[t]
\centering
\includegraphics[scale=0.33]{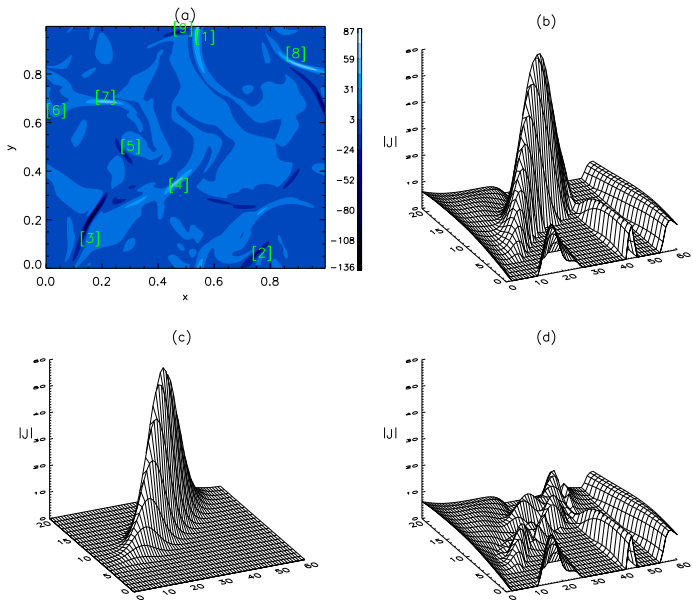}
\caption{Example of current sheet identification and fitting. } 
\label{csfit}
\end{figure}

Our approach consists of two steps. First, an ad-hoc
thresholding algorithm identifies current sheet candidates by simply taking all pixels 
in $|J|$ above a pre-defined fraction of $|J|_{max}$ and testing for contiguity of 
the selected regions. This is done in two-dimensional cross-sections of the loop
simulations, with the algorithm accounting for the periodic boundary conditions 
used by the pseudo-spectral RMHD scheme. By this we mean that
a current sheet that appears at a border of the simulation box 
will appear at the other border (or at up to 4 edges if it appears at a
corner), but will be identified only one occurrence. This
feature is crucial, considering that we are automating this procedure to analyze
tens of thousands of simulation cube samples and the likelihood of current sheets
appearing at domain edges is quite high. Figure~\ref{csfit}(a) shows a contour
plot of current density for a cross-section of one of our simulations. 
Current sheet candidates identified by the routine are
labeled by green bracketed numbers. The structures labeled $[1]$ and $[6]$, for
example, appear at edges but are uniquely identified. 

After current sheet candidates are identified they are morphologically examined 
by another automated algorithm, which performs least-square fitting with a
bi-variate Gaussian. The automated algorithm is implemented using fitting and 
parameter constraining tools found in the Package for the Interactive Analysis of 
Line Emission (PINTofALE, \citealt{PoA2000}). 

Together, these two algorithms yield current sheet orientations with
respect to the axes ($\theta$), number of current sheets present ($N$), local
$J_{max}$, together with $\sigma_{small}$ and $\sigma_{large}$, which serve as
proxies for current sheet width ($\lambda$) and current sheet length ($\Delta$),
respectively. The three other panels in Figure~\ref{csfit} show one such fit
for the current sheet labeled [9]. A surface plot shows $|J|$ in the region where current sheet [9]
resides in plot (b). Plot(c) shows the bi-variate Gaussian fit and (d) shows
the residual. 
The primary shortfalls of this approach can be summarized as follows: 
$(1)$ Many current sheets are not well approximated by bi-variate Gaussians. Profiles are
often asymmetric, and the 2-D support of the current sheet structures is often
bow-shaped rather than linear. 
$(2)$ Because we are taking only discrete samples in time  (full data cubes are
saved at a pre-determined cadence during simulations runs), the measurements
will be biased towards current sheet structure that is most long-lived during
the lifetime of the sheets. 
$(3)$ In the present analysis, the threshold for structure detection
is set at $10\%$ of maximum $|J|$ (as measured in each time step), low enough
so that most local maxima are included. Unfortunately, this renders the
iterative search approach we take, while robust, quite computationally expensive. 
At higher resolutions, this becomes prohibitive, and requires re-sampling to
lower resolution for reasonable run-times.

\begin{figure*}[t]
\centering
\includegraphics[scale=0.33]{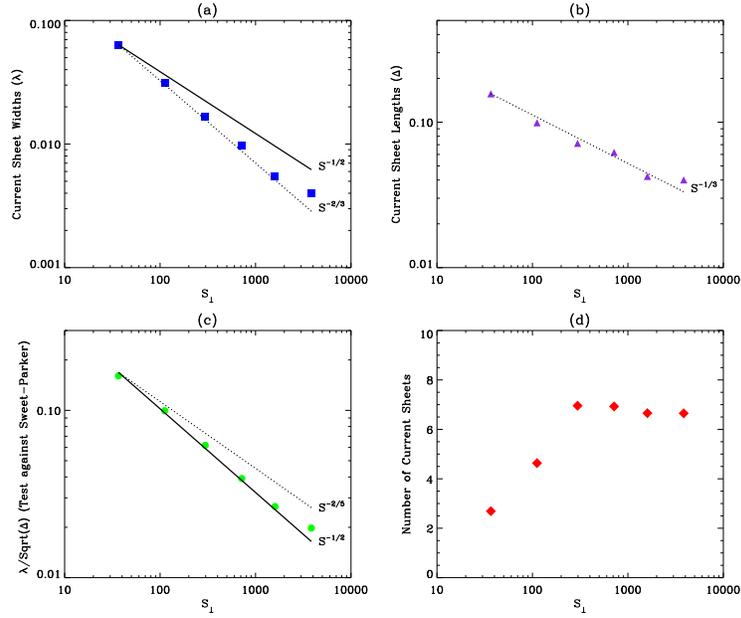}
\caption{Scaling of measured current sheet parameters with Lundquist number.}
\label{cstat}

\end{figure*}

\section{Summary of Results \& Conclusions} 
\label{4conc}

Our current sheet identification and fitting routines have been
applied to high resolution data set of
\cite{nlb2012}, producing a large sample of fits 
in the range of thousands to tens of thousands of current sheets per 
simulation run. 

In Figure~\ref{cstat}, we report weighted average quantities,
where we use goodness-of-fit from the least square bi-variate Gaussian fitting
as the weighting factor. Plots (a) and (b) show weighted average values for
current sheet widths and lengths as a function of Lundquist number
$S_{\perp}$. Plot (c) demonstrates good agreement with Sweet-Parker scaling.
Plot (d) shows the number of current sheets averaged over all post-processed
time slices. These results provide some empirical support for the
assumptions we used for our scaling analysis. It is noted however that, even
if the Sweet-Parker scaling of $\lambda/\Delta^{-1/2} \propto S_{\perp}^{-1/2}$ is
recovered, both the current sheet width ($\lambda \propto S_{\perp}^{-2/3}$) 
and current sheet length ($\Delta \propto S_{\perp}^{-1/3}$) decrease with the
increase of $S_{\perp}$ faster than the
scalings ($\lambda \propto S_{\perp}^{-1/2}$, $\Delta \sim$ constant) 
assumed in both \cite{nlb2012} and \cite{ls1994}. 

A more detailed analysis of these current sheet fitting results, and an
extension of this analysis to examine the 3D structure of current layers
is now underway. Of particular interest here is how the spatial
separation of dissipative events in these simulations can inform an
analysis of flare energy distributions, which typically only consider temporal
variations in event definition (cf. \citealt{Buchlin2005, nl2012}). 

\acknowledgements 
Computer time was provided by UNH (using the Zaphod 
cluster at the Institute for the Study of Earth, Oceans, and Space), and 
HPC resources from the Arctic Region Supercomputing Center, the
DoD High Performance Computing Modernization Program.
This research is supported in part by grants from  NSF
(AGS-0962477 and AGS-0962698) and NASA (NNX08BA71G, NNX09AJ86G, NNX06AC19G,
and NNX10AC04G) and DOE (DE-FG0207ER46372),
and by NSF through TeraGrid resources provided by NCSA under grant number TG-PHY100057 
and through computational resources provided by NICS under grant UT-NTNL0092.

\bibliographystyle{asp2010}
\bibliography{thesis}

\end{document}